\documentstyle[aps]{revtex}
\begin{document}
\draft
\title{Monte Carlo Study of Si(111) homoepitaxy}
\author{Makoto Itoh}
\address{Interdisciplinary Research Centre for Semiconductor Materials, \\
 The Blackett Laboratory, Imperial College, London SW7 2BZ, 
 United Kingdom}
\maketitle
\begin{abstract}
  An attempt is made to simulate the homoepitaxial growth of a Si(111) surface 
by the kinetic Monte Carlo method in which the standard Solid-on-Solid model 
and the planar model of the $(7\times7)$ surface reconstruction are used in 
combination. 
  By taking account of surface reconstructions as well as atomic deposition 
and migrations, it is shown that the effect of a coorparative stacking 
transformation is necessary for a layer growth. 
\end{abstract}

In contrast to a number of experimental studies on a Si(111) homoepitaxy\cite{cerullo90,koehler89,koehler91,ichimiya93,tochi93,naka91B,shima94,hasegawa94,shige95,hoshino95,ichimiya96A,ichimiya96B,ichimiya96C,voigt96A,naka91A,shige96A}, 
there are no theoretical studies on it, obviously because of the great 
complexity of its structure known as the dimer-adatom-stacking-fault (DAS) 
reconstruction\cite{taka85}. 
Indeed, scanning tunneling microscopy (STM) observations have been the only 
accessible method to elucidate the growth behavior at the atomic level. 

In the previous papers, I have proposed the four-state planar model of the 
Si(111) surface reconstruction, in which the three-point discrete planar 
rotators (3PDR) were introduced to denote the atoms belonging to the lower 
half of the topmost bilayer (BL). 
Then, by doing the calculations with the state-flipping dynamics, I have 
investigated the properties of the $(7\times7)$-to-``($1\times1$)'' phase 
transition\cite{itoh96}. 
Here, the same model is used to examine the growth behavior of a 
Si(111) homoepitaxy by introducing hopping events as well. 
This is accomplished by combining the 3PDR model with the Solid-On-Solid 
(SOS) model\cite{clarke87} and let the latter undertake the center 
force part of the interactions. 

In addition to the four states, {\it i.e.} unfaulted-stacking (US) states, 
faulted-stacking (FS) states, and two kinds of states which constitute dimer 
pairs (DA and DB) in Ref.\ \cite{itoh96}, I am going to 
introduce a bulk (BK) state, a vacant (VC) state, and an intermediate (IM) 
state which deposited atoms are supposed to take before incorporated into 
bilayer structures. 
Actually, its importance to dissolve dimer rows and FS halves of the DAS 
structures were pointed out experimentally in relation to the island 
nucleation and the step flow dynamics\cite{koehler89,koehler91,ichimiya93,tochi93,naka91B,shima94,hasegawa94}. 
Evidently, for these dynamics to be realized, the dissociation of FS halves 
and dimer rows must proceed simultaneously, and thus atoms in the IM state 
must change their states in a coorparative fashion\cite{ichimiya93}. 
To accomplish this, I will choose the atoms belonging to the lower 
halves of the BL's of a Si(111) surface as the ingredients of the model, and 
apply the combined model for which the algorithm a la Maksym\cite{maks88} 
is employed. 
Besides, I will assume the FS states to exist only on the topmost BL, 
and prohibit the occurrence of them on underlayers. 

In brief, all terms in the model are described in terms of the 3PDR's states, 
the basis vectors ${\bf a}_i$ with $i =0,1,2$ on the 
two-dimensional lattice $\Lambda$ shown in Figs.\ \ref{fig:terms} (a--b). 
The first term is, 
\begin{eqnarray}
H_0 \; = - J_0 
\sum_{{\bf x},{\bf x'} \in \Lambda}\; 
\biggl(
& \sum_{\zeta \in \{ US, FS \} } &
\delta_{\varphi({\bf x}), \varphi({\bf x'})}\; \delta_{\varphi({\bf x}),
\zeta}\;
-\; \delta_{\varphi({\bf x}), {\rm US}}\;
\delta_{\varphi({\bf x'}), {\rm FS}}\;  \nonumber\\
&+&\; \delta_{\varphi({\bf x}), {\rm US}}\; 
\delta_{\varphi({\bf x'}), {\rm BK}}\; 
-\; \delta_{\varphi({\bf x}), {\rm FS}}\;
\delta_{\varphi({\bf x'}), {\rm BK}}\;
\biggr) \; 
,\label{eq:order}
\end{eqnarray} 
which generates the US-ordered states, the FS-ordered states, and the bulk 
structures as shown in Figs.\ \ref{fig:terms} (c--d). 
In this equation, the ${\bf x}$ and ${\bf x'}$ summations are taken over 
all the nearest-neighbor pair sites on $\Lambda$. 
The symbol $\varphi({\bf x})$ denotes the state of a 3PDR at site ${\bf x}$ 
on $\Lambda$  and $\delta_{\varphi,\varphi'}$ denotes the Kronecker's delta. 

The dimer-adatom interaction (DAI) 
(Refs.\cite{stich92,payne87,itoh93}) 
terms and those connecting them are drawn in 
Figs.\ \ref{fig:terms} (e--h), which are realized respectively by 
the three-point interactions as 
\begin{eqnarray}
H_1 = - & J_1 &
\sum_{{\bf x} \in \Lambda}
\sum_{i \in Z_{\rm 3}}\;
\delta_{\varphi({\bf x}), {\rm DA}}
\delta_{\varphi({\bf x} + {\bf a}_i),{\rm DB}}\; \nonumber\\ &\times&
\,\left[
  \; \delta_{\varphi({\bf x}-{\bf a}_{i+1}), {\rm US}} 
+ \; \delta_{\varphi({\bf x}-{\bf a}_{i+2}), {\rm FS}}\;
\right]\,
,\label{eq:DAI}
\end{eqnarray} 
and the terms obtained by interchanging the symbols DA and DB in 
Eq.~(\ref{eq:DAI}). 
Here, the summation over ${\bf x}$ is over single sites on $\Lambda$, while 
the equivalence ${\bf a}_i = {\bf a}_j$ holds for the basis vectors 
such that $i\equiv j \pmod{3}$ in the summation over the crystallographic 
directions $(i \in Z_{\rm 3})$. 
The fact that BK states do not appear in these summations means that 
when an atom sticks onto a US site which is interacting 
with adjacent dimers, the DAI's are diminished or even lost, as in reality 
the dimer contraction\cite{stich92} takes place only when adatoms are 
associated with dimers. 

The small stacking energy difference between FS and US sites is defined by 
adding $\pm J_3$ to each site, respectively. 

Since I will consider hopping events as well as state-flipping ones, corner 
holes in the DAS structure must be properly dealt with, as opposed to the 
previous studies in which the center atoms of corner holes remained on a 
lattice\cite{itoh96}. 
Thus, the interaction term has to be modified to make the central site of a 
corner hole vacant, which is achieved by 
\begin{eqnarray}
H_4 = - & J_4 &
\sum_{{\bf x}\in \Lambda}
\sum_{i\in Z_3}
\biggl( 
\delta_{\varphi({\bf x}), {\rm DA}}
\delta_{\varphi({\bf x} + {\bf a}_i),{\rm DB}}\; \nonumber\\ &\times&
\,\left[
  \delta_{\varphi({\bf x} - {\bf a}_{i+1}), {\bf VC}}\; 
  \delta_{\varphi({\bf x} - {\bf a}_{i+2}), {\bf US}}\; 
+\; \delta_{\varphi({\bf x} - {\bf a}_{i+1}), {\bf FS}}\; 
    \delta_{\varphi({\bf x} - {\bf a}_{i+2}), {\bf VC}}\; 
\right]\,
\; \nonumber\\
 & + &
\delta_{\varphi({\bf x}), {\rm DB}}
\delta_{\varphi({\bf x} - {\bf a}_i),{\rm DA}}\; \nonumber\\ &\times&
\,\left[
  \delta_{\varphi({\bf x} + {\bf a}_{i+1}), {\bf VC}}\; 
  \delta_{\varphi({\bf x} + {\bf a}_{i+2}), {\bf FS}}\; 
+\; \delta_{\varphi({\bf x} + {\bf a}_{i+1}), {\bf US}}\; 
    \delta_{\varphi({\bf x} + {\bf a}_{i+2}), {\bf VC}}\; 
\right]\,
\biggr) \; 
.\label{eq:corner} 
\end{eqnarray}
The diagrammatic representations of them are shown in 
Figs.\ \ref{fig:terms} (i--j), in which the cross symbols denote the 
vacancies. 

Finally, I will introduce an additional term to compensate for unexpected 
contributions which may arise from the SOS model part, because the center 
force tends to fill in the central site of a corner hole with an unnecessary 
atom. 
This compensation term is given by, 
$
H_5 =  2 J_4 
\sum_{{\bf x} \in \Lambda}
\sum_{i \in Z_{\rm 3}}\;
\,\left(
  \delta_{\varphi({\bf x}), {\rm US}}
+ \delta_{\varphi({\bf x}),{\rm FS}}\; 
\right)\,
\delta_{\varphi({\bf x}+{\bf a}_{i}), {\rm DA}} 
\delta_{\varphi({\bf x}-{\bf a}_{i}), {\rm DB}}\;
$.

Thus, the total Hamiltonian is given by the sum of the terms explained above 
(3PDR part) plus the SOS part as $H_{tot} = H_{SOS} + \kappa  H_{3PDR}$, 
where $\kappa$ denotes their relative weight.
Correspondingly, the kinetic barrier for the growth simulation is defined 
by $- {\cal H}_{tot}({\bf x})$  if $ {\cal H}_{tot}({\bf x}) < 0 $ and 
zero otherwise. 
Here, ${\cal H}_{tot}({\bf x})$ is defined from $H_{tot}$ 
by $H_{tot} = \sum_{{\bf x} \in \Lambda} {\cal H}_{tot}$. 
Then, I will consider as kinetic events the deposition of atoms, 
a state-flipping event, and a hopping of an atom followed by a flipping of 
its own state. 

By the STM observations, the back-bonding energy  of silicon atoms on a 
Si(111) surface was estimated to be $1.020\; eV$\cite{voigt96A}. 
However, since the atoms belonging to the upper half of a BL are also 
implicitly taken into account in the model, it is necessary to include their 
effect on the back-bonding energy of the SOS part $J_{sub}$ 
as well to give it a slightly larger value. 
On the other hand, Ichimiya and co-workers estimated the activation energies 
of the island decay and hole-filling phenomena to be around $1.5\; eV$ 
and $1.3\; eV$, respectively\cite{ichimiya96A,ichimiya96B}. 
As opposed to the former case, these values should be larger than $J_{sub}$ 
because other interaction effects are included in their measurements. 
Therefore, I will take $J_{sub} = 1.20 \;eV$, which is near the average 
value of the reported ones. 

At high temperatures, surface structure formation will be dominated by the SOS 
part, and thus the coupling constant for the lateral part $J_{lat}$ will give 
the principal contribution to the surface melting temperature $\approx$ 1400K. 
Therefore, it is natural to set $J_{lat} = 0.12\; eV$. 

As for the 3PDR part, the ratios $J_{i}/J_0 \; (i = 1,\ldots,4)$ of the 
coefficients of the interaction terms $H_{i}$ appropriate in producing 
the $(7\times7)$ DAS structure were already known\cite{itoh96}. 
To be precise, $J_4$ has not yet been estimated, because the $H_4$ term in 
Eq.~(\ref{eq:corner}) takes the different form than in Ref.\ \cite{itoh96}. 
However, this difference arises merely due to the absence of hopping 
processes in the former work, and the same value for $J_4$ should be 
applicable also to the present case. 
The ratios of the parameters thus used in the calculations are given by, 
$J_1/J_0 = 4.175, \; J_2/J_0 = 2.450, \; J_3/J_0 = 0.100$, 
and $J_4/J_0 = 0.500$. 

Now, the only remaining unfixed parameter is $\kappa$. 
In the previous study\cite{itoh96}, it was also shown that the transition 
temperature $T_{t}$ of the $(7\times7)$ reconstruction was given approximately 
by $J_0$, so that $\kappa J_0 \approx k_{B} T_{t}$,  
where $k_{B}$ is the Boltzmann's constant. 
Then, since $T_{t}$ has been reported to be about 1100K\cite{bauer85}, I will 
set $\kappa J_0 = k_{B} T_{t} = 9.5\times10^{-2} \; eV$. 

It has been shown experimentally that the complexity and the stability of 
the $(7\times7)$ structure obstruct the growth. 
Moreover, when atoms are deposited, the kinetic process has been shown to 
proceed by first forming the clusters of IM states and later by their 
transformation into US states\cite{cerullo90,koehler89,koehler91,ichimiya93,tochi93,naka91B,shige95}.  
To accomplish this, I will introduce the interaction energies of this state, 
{\it i.e.} I will denote the lateral interaction between a pair of IM states 
by $E_{lat}^{*}$, and its back-bonding energy on a US site by $E_{US}^{*}$ 
and that on a FS site by $E_{FS}^{*}$, respectively. 

Suppose that an atom is going to stick onto a site ${\bf x}$ with its three 
substrate atoms in a plaquette 
at ${\bf x}$, ${\bf x}+{\bf a}_{2}$, and ${\bf x}-{\bf a}_{1}$. 
Then, the new state of an atom after a deposition or a hopping is determined 
by the following rules: 
(1) if all three substrate atoms are in the US states, 
then set it to be in the US state if any one of the adjacent atoms in the 
same plane is in the US or in the bulk state, and if not, set it to be in the 
IM state if any one of the adjacent atoms in the same plane is in the IM 
state or it is isolated, 
(2) if some of the substrate atoms are either in the US states or in the FS 
states but both are not simultaneously present, then set it to be in the IM 
state if any one of the adjacent atoms in the same plane is in the IM state, 
(3) if all three substrate atoms are in the FS states, set it to be in the 
IM state, 
and 
(4)If none of them is met, an atom is prohibited to stick there. 
Here, due to the bilayer character of the 3PDR's, I will assume no bonding is 
available for an IM-state atom to prevent further sticking of an 
atom on it. 
If the size of an IM-state cluster exceeds a critical size $S_{cr}$, 
a stacking transformation takes place and all atoms in the cluster is set to 
be in the US states.

The length of the random hopping process is set to be 50, so 
that the effective radius of the searching area approximately becomes equal 
to the linear dimension of the $(7\times7)$ unit cell. 
Then, by carrying out the calculations, I found it appropriate to 
choose $E_{lat}^{*} = 1.35 \; eV$, $E_{US}^{*} = 1.20 \; eV$ 
and $E_{FS}^{*} = 1.30 \; eV$ to produce an oscillatory behavior for the  
step density (SD) when the atomic flux and the substrate temperature are 
chosen to be 0.1 monolayer coverage per second and 650 K, respectively. 
In Fig.\ \ref{fig:SD}, (1 - SD) is plotted as a function of a coverage 
for which the hopping length is 50, and $S_{cr} = 17$ is used. 
The lattice sizes are $35\times35$ and $49\times49$, where the snapshots 
of the latter are shown in Fig.\ \ref{fig:snap}.

Interestingly, the oscillatory behavior of the SD was seen 
only when $\kappa J_0$ is very close to $9.5\times10^{-2} \; eV$. 
Unfortunately, however, this does not look like the ordinary BL-by-BL growth 
behavior observed by reflection high-energy electron diffraction (RHEED) 
experiments and by STM observations\cite{shige95,naka91A,shige96A}. 
Instead, it looks similar to the RHEED intensity oscillation obtained for 
a Ge/Si(100) heteroepitaxy\cite{xie}. 
This means that atop sites of a growing surface is favored for further growth. 
At the same time, this island-growth behavior appeared due partly to the 
limitation of the 3PDR model, with which only the bilayer character of the 
growth can be realized, and hence this discrepancy implies the true IM state 
does not possess this character. 
In other words, with the present model, there is no guarantee that the 
stacking transformation into the US ones is irreversible, and 
this property seems to play a very crucial role in the surface growth. 
Actually, in the present model, it is not possible to preclude an event in 
which an atom in a US state to hop around and stick to an IM island and change 
its state into an IM one. 

In summary, by the combined use of the 3PDR model\cite{itoh96} and the SOS 
model\cite{clarke87}, the kinetic growth simulation of Si(111) homoepitaxy 
is performed. 
It is found that atop sites of a growing surface is favored for further 
growth. 
Also, it is found that for the bilayer step density to show the oscillatory 
behavior, it is necessary to choose the back-bonding energies so that 
the inequality $J_{sub} = E_{US}^{*} < E_{FS}^{*}$ holds. 
However, this result is evidently related to the bilayer character of 
the 3PDR model. 
More importantly, the irreversibility of the transformation from an 
IM state to a US one is crucial in obtaining the BL-by-BL growth 
mode. 

\acknowledgments
The author would like to thank K.\ Takayanagi for helpful discussions, 
H.\ Fujino and Y.\ Shigeta for sending him their articles, 
and M.\ Blencowe and D.\ D.\ Vvedensky for reading through the manuscript.

\begin{figure}
  \caption{
  Interaction terms of the Hamiltonian are displayed diagrammatically. 
  (a) Allowed four states of the 3PDR's, US,FS,DA, and DB. 
  (b) Three basis vectors used in the equations. 
  (c--d) Locally US and FS ordered states. 
  (e--f) DAI terms. 
  (g--h) DAI-connecting terms. 
  (i--j) Corner-hole stabilizing terms. 
  }\label{fig:terms}
\end{figure}

\begin{figure}
  \caption{(1 - SD) plotted against the coverage. 
  The lattice sizes are $35\times35$ and $49\times49$, 
  the hopping length is 50, and the critical cluster size is 17. 
  }\label{fig:SD}
\end{figure}

\begin{figure}
  \caption{Snapshots of simulation on Si(111) homoepitaxy. 
  The lattice size is $49\times49$. 
  Atoms at higher layers are depicted by darker and larger symbols, where 
  3PDR's denote US,FS,DA, and DB states, and shaded discs denote IM states. 
  (a) 0.40 BL.
  (b) 0.70 BL.
  (c) 1.00 BL. 
  (d) 1.50 BL.
  }\label{fig:snap}
\end{figure}

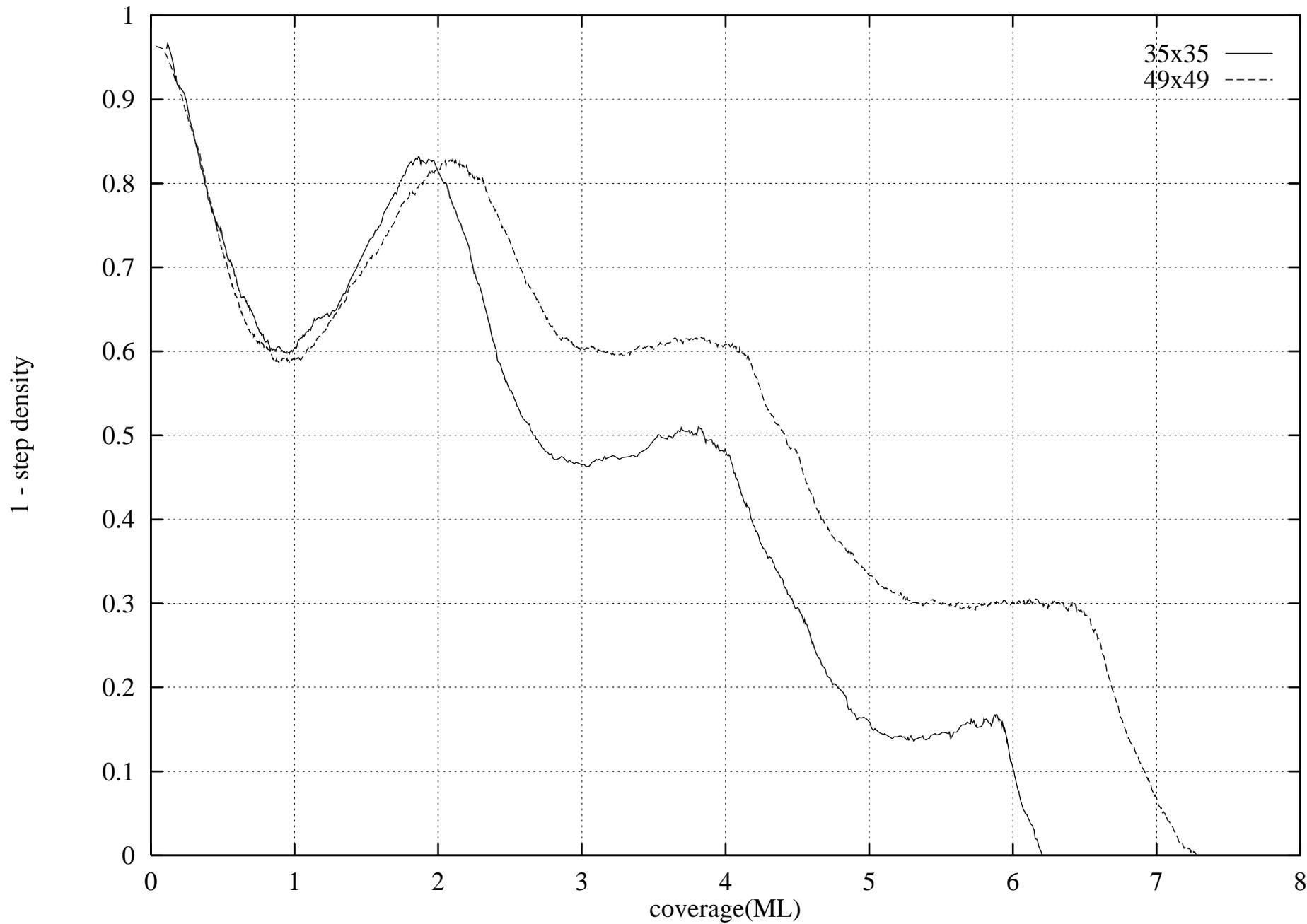
\end{document}